\documentclass[aps,prd,fleqn,superscriptaddress,twocolumn]{revtex4}
\usepackage{graphicx,color,natbib}
\usepackage{amsmath,amssymb,amsfonts}

\usepackage{epsfig,epsf}
\usepackage{dcolumn}
\usepackage{float}
\usepackage{bm}

\newcommand{\bse}{\begin{subequations}}
\newcommand{\ese}{\end{subequations}}
\newcommand{\be}{\begin{equation}}
\newcommand{\ee}{\end{equation}}
\newcommand{\bea}{\begin{eqnarray}}
\newcommand{\eea}{\end{eqnarray}}
\newcommand{\ba}{\begin{array}}
\newcommand{\ea}{\end{array}}

\input amssym.def
\input amssym.tex

\usepackage[colorlinks=true, linkcolor=blue, bookmarks=true]
{hyperref}

\begin{document}

\title{Two Concepts of Holographic Complexity under Thermal and Electromagnetic Quenches}

\author{Mohammad Ali-Akbari\footnote{m\_aliakbari@sbu.ac.ir}}
\affiliation{Department of Physics, Shahid Beheshti University, 1983969411, Tehran, Iran
}
\author{Mahsa Lezgi\footnote{s\_lezgi@sbu.ac.ir}}
\affiliation{Department of Physics, Shahid Beheshti University, 1983969411, Tehran, Iran
}

\begin{abstract}
We study the evolution of holographic subregion complexity (HSC) in a thermally and magnetically quenched strongly coupled quantum field theory in $2+1$ dimension. We illustrate two concepts of complexity in this theory, (1): how much information it takes to specify a state     by studying the behavior of the final value of HSC in terms of the final temperature and magnetic field  and (2): how long it takes to reach the state, by considering the time it takes for HSC to relax as a function of the final temperature and magnetic field. 
In the first concept, we observe that the effect of temperature and magnetic field on HSC is decreasing until the energy of the probe is comparable to the final temperature and magnetic field. We present an argue based on an ensemble of microstates corresponding to a given mixed macrostate. In the second concept, we show that the time of relaxation of HSC decreases with the increase of temperature and magnetic field for fixed value of the energy of the probe.
We also compare the time evolution of HSC for two quenches, in the first concept. We observe that the absolute value of the ratio of the final value of HSC for two kinds of quenches depends on the energy of the probe.
\end{abstract}
\maketitle
\tableofcontents

\section{Introduction}

Studying the evolution of out-of-equilibrium systems as they move towards equilibrium is a crucial aspect of physics. When these systems are both far from equilibrium and strongly coupled, they become considerably more complex to investigate. To address such non-perturbative problems, including far-from-equilibrium phenomena, one of the tools that physicists have employed is the gauge-gravity duality, or the holographic idea. This duality as a useful framework maps strongly coupled field theories to weakly coupled gravity in one more dimension \cite{CasalderreySolana:2011us}. A noteworthy example of a strongly coupled and far-from-equilibrium medium is the Quark-gluon plasma (QGP) generated during heavy ion collisions at the Relativistic Heavy Ion Collider (RHIC). Experimental observations have revealed that the time scale required for this plasma to reach thermal equilibrium is significantly shorter than what perturbative methods would predict \cite{one}. 

Such thermalization processes are discussed in the subject of holographic thermalization. One of the ways that a far-from-equilibrium state is prepared is that a limited duration source in time is activated in the boundary field theory. Prior to activating the source, the system exists in its ground sate. The external source performs work on the system, causing it to reach an excited state. This excited state undergoes temporal evolution and eventually, upon deactivating the source, settles into a thermal equilibrium state. The representation of this process on the gravity side is encoded in the process of the formation of a black hole in the bulk \cite{bala}. A thermal quench in the boundary theory corresponds to the collapsing of a shell of uncharged matter to Anti-de Sitter (AdS) space-time and formation of a Schwarzschild black hole. An electromagnetic quench can be dual to the addition of a shell of charged matter to AdS space-time and formation of extremal dyonic black hole \cite{metric2}, as reviewed in this paper. Time evolution of this far-from-equilibrium system can be probed by local and non-local observation and the HSC is the probe we are interested in here.

The holographic concept establishes a connection between quantum information theory quantities and specific geometric quantities in the bulk theory. For instance, the Hubney-Rangamani-Takayanagi (HRT) proposal, which utilizes entanglement entropy as a measure of quantum correlation in a pure quantum state, serves as a straightforward geometric approach \cite{mm1} that has passed numerous tests successfully \cite{Takayanagi,mm2,mm3,mm4,mm5,mm6,mm7,mm8,mm9}. Complexity, a fundamental concept in quantum information theory, is defined as the minimum number of simple gates required to generate a given state from a reference state \cite{john}. This quantity refers to the time and space resources needed to perform a computation efficiently \cite{coco1}. In the realm of quantum field theory, complexity pertains to the minimum number of unitary operators needed to transform a reference state into a target state \cite{circuit}. Essentially, complexity can be used to classify different quantum states based on the difficulty of their preparation. Within the holographic framework, two conjectures have been proposed to describe complexity: the CV (complexity=volume) conjecture and the CA (complexity=Action) conjecture. The complexity in the CA conjecture is determined by evaluating the bulk action on the Wheeler-de Witt patch, which is anchored at a specific boundary time. On the other hand, in the CV conjecture, complexity is defined as the volume of a codimensional-one hypersurface in the bulk that terminates on a time slice of the boundary \cite{susskind1,susskind2}. Initially introduced for the complexity of pure states in the entire boundary system, both conjectures can be extended to encompass the complexity of mixed states in corresponding subregions \cite{comments,alishahiha}. 
Inspired by the Hubney-Ryu-Takayanagi proposal, the CV proposal extends to encompass subregions, corresponding to the complexity of mixed states, and is referred to as HSC, in which the complexity of a subsystem on the boundary is determined by the volume of codimensional-one hypersurface enclosed by  Hubney-Ryu-Takayanagi surface \cite{alishahiha}. Numerous studies exploring the CV and CA conjectures, as well as HSC for various gravity models, can be found in the literature \cite{volume,zhang1,zhang2,renormalization,faregh,
mozaffar,mozaffar2,sub,asadi,mahsa,m1,m2,m3}.

In this paper we study the evolution of HSC on a far-from-equilibrium state, under a thermal and an electromagnetic quench for a 2+1 strongly coupled quantum field theory. We numerically calculate the final value of HSC and the relaxation time, the time it takes for HSC to reach a constant value, as the the system moves towards the equilibrium. We focused on the effect of the energy of the probe, which is characterized with the length of the subregion considered in the boundary as the given mixed state, on the behavior of HSC.
Our idea is that in the study of quenched systems holographically, we can highlight two distinct concepts of complexity that is introduced in quantum information literature \citep{coco1}. That is, how much information it takes to specify a state and how many operations are required to reach the state. The final value of HSC as a function of temperature and magnetic field can represent the first concept and the relaxation time may indicate the second one.
We start with a short review on the backgrounds in section \ref{21} and then compute the HSC in section \ref{211} and discuss its behavior in numerical results section \ref{2111}. 

\section{Review on the background}\label{21}

In this section we review the time dependent asymptotically $\rm{AdS}_{4}$ geometries corresponding to a thermal and electromagnetic quenched quantum field theory. Holographically, a quench in a quantum field theory can be represented as the collapse of a thin shell of null dust falling from the $\rm{AdS}$ boundary to form a black hole. This phenomenon can be effectively described using Vaidya-$AdS$ metric. The metric of the Vaidya-$\rm{AdS}_{4}$ can be written as follows \cite{bala},
\begin{align}
ds^{2}=\frac{1}{z^{2}}\left(-f(z,v)dv^{2}-2dzdv+d\vec{x}^2\right).
\label{metr1}
\end{align}
The coordinate $v$ labels the path of the ingoing null trajectory and aligns with the time coordinate $t$ on the boundary as $z$ approaches zero. $z$ is the radial coordinate, $\vec{x}\equiv(x_1,x_2)$ and
the $\rm{AdS}$ radius is rescaled to be unit. For the thermal quench in which collapsing of a uncharged source leads to the formation of a $\rm{AdS}_{4}$-Schwarzschild black hole at late times, corresponding to a thermal state in the dual boundary quantum filed theory, we have
\begin{align}
f(z,v)=1-\left(\frac{z}{z_{0}(v)}\right)^{3},~~~~~~~~~m(v)=z_{0}(v)^{-3},
\label{metr2}
\end{align}
where $m(v)$ is the mass function of the in-falling shell. The subsequent functional form is considered for $z_{0}(v)$ \cite{metric1},
\begin{align}
z_{0}(v)=z_{\infty}\left(\frac{1+\tanh(v/v_{0})}{2} \right)^{-1/3},
\label{metr3}
\end{align}
in which $v_{0}$ characterizes the thickness of the shell or the duration of the quench. The parameter $z_{\infty}$ can be understood as being connected to the final temperature of the background in the late time regime \citep{metric1},
\begin{align}
T=\frac{3}{4\pi}z_{\infty}^{-1}.
\label{metr4}
\end{align}
It is possible to study a completely different type of quench in this system. This non-thermal quench is achieved by introducing electric and magnetic sources in the theory. To ensure a fully non-thermal system, we approach the extremal (zero temperature) black hole solution at late times \cite{metric1}. The electric components at late times correspond to non-zero charge density and chemical potential in the dual 2+1 dimensional boundary quantum field theory \cite{metr3}, while the magnetic components correspond to a background magnetic field \cite{metr4}. Considering the presence of the electric-magnetic duality invariance of the four dimensional gravity system, different choices of electric or magnetic sources can be interchanged, and for convenience, it is often considered that the system are undergoing a purely magnetic quench. In this case, the function $f$ is taken to be \cite{metric1},
\begin{align}
f(z,v)=1-4\left(\frac{z}{z_{0}(v)}\right)^{3}+3\left(\frac{z}{z_{0}(v)}\right)^{4}.
\label{metr5}
\end{align}
The same quench profile for $z_{0}(v)$, the equation \eqref{metr3}, is chosen and this time the parameter $z_{\infty}$ determines the final strength of the magnetic field  \cite{metric1}, 
\begin{align}
B=\sqrt{3}z_{\infty}^{-2}.
\label{metr6}
\end{align}

We use the metric \eqref{metr1} with \eqref{metr2} and \eqref{metr5} for the thermal and electromagnetic quench, respectively, to discuss time evolution of the HSC. In order to further study these backgrounds and examine holographic entanglement entropy in them, refer to \cite{metric1}.

\section{Holographic subregion complexity}\label{211}

\begin{figure}[ht]
\centering 
\includegraphics[width=70 mm]{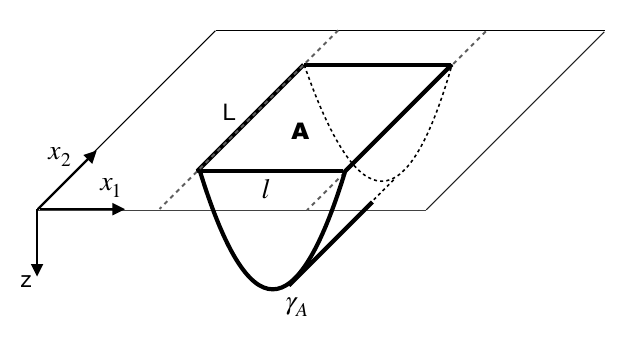}
\caption{A strip entangling surface of length $l$ and width $L \rightarrow \infty$ in a static $\rm{AdS}$ geometry.}
\label{fig0}
\end{figure}

The mixed state's complexity, which corresponds to a subregion $A$ on the boundary, is linked to the volume contained by the extremal surface $\gamma_{A}$ that appears in the computing of holographic entanglement entropy by the HRT proposal \cite{alishahiha}, i.e.
\be 
{\cal{C}}_{A}=\frac{V_{\gamma_{A}}}{8\pi R G_{N}},
\label{cv1}
\ee
where $R$, $G_{N}$ and ${\cal{C}}_{A}$ are $\rm{AdS}$ radius, Newton constant and the HSC for the subregion $A$, respectively. To calculate $\gamma_{A}$ at a given time, we take into account subregion $A$, which has been defined as,
\begin{align}
A:=x_{1}(\equiv x)\in \left(-\frac{l}{2},\frac{l}{2}\right)~~~,~~~x_{2}\in\left(-\frac{L}{2},\frac{L}{2}\right),
\label{cv2}
\end{align}
which describes a strip surface of finite length $l$ and width $L \rightarrow \infty$ as shown in figure \ref{fig0} for a static $\rm{AdS}$ background. In general, $\gamma_{A}$ do not live on a constant slice for a dynamical background. Due to the symmetry of the strip, it is possible to represent $\gamma_{A}$ in the bulk using this parametrization,
\be
v=v(x),\ z=z(x),\ v(\pm\frac{l}{2})=t-\epsilon,\ z(\pm\frac{l}{2})=\epsilon,
\label{cv3}
\ee
where $\epsilon$ is a $\rm{UV}$ cut-off. Then the area of the minimal surface is obtained easily using \eqref{metr1},
\begin{align}
S=\frac{L}{4G_{N}}\int_{-\frac{l}{2}}^{\frac{l}{2}}\frac{\sqrt{1-f(z,v)v'^{2}-2z'v'}}{z^{2}}dx.
\label{cv4}
\end{align}
We treat the integrand in equation \eqref{cv4} as a Lagrangian and the symmetry of the strip causes that the turning point of the extremal surface, $\gamma_{A}$, locating at $x=0$, in which,
\be\label{cv5}
v'(0)=z'(0)=0,\ v(0)=v_*,\ z(0)=z_*.
\ee
The equations of motion for $z(x)$ and $v(x)$ can be determined, and subsequently, by employing \eqref{cv5}, they can be solved numerically to obtain the profiles $z(x)$ and $v(x)$. The volume can be parameterized by $v=v(x)$ and $z=z(x)$, or equivalently $z=z(v)$. Therefore, the volume for the background solution \eqref{metr1}, by choosing an appropriate function $f$, \eqref{metr2} and \eqref{metr5} for the thermal and electromagnetic quenches respectively, becomes
\begin{align}
V=2L\int_{v_{*}}^{v(\frac{l}{2})}\left(-f(z(v),v)-2\frac{\partial z}{\partial v}\right)^{\frac{1}{2}}z(v)^{-3}x(v)dv.
\label{cv6}
\end{align}
Since the HSC is divergent, it is convenient to consider subtracted HSC (a normalized version of HSC) using \eqref{cv1}, as follows
\begin{align}
C\equiv\frac{8\pi RG_{N}(\mathcal{C}-\mathcal{C}_{AdS})}{L}=\frac{V-V_{AdS}}{L},
\end{align}
where $\mathcal{C}$ and $\mathcal{C}_{AdS}$ are the HSC for $A$ in \eqref{metr1} and $\rm{AdS}$ geometry respectively. The volumes are defined with respect to the same boundary region such that $V$ in equation \eqref{cv6} reduces to $V_{AdS}$ by setting $f$ equal to one.

To determine the relaxation time for HSC, we introduce the following function
\be
{\epsilon}(t)=|1-\frac{C(t)}{C(\infty)}|.
\ee
The relaxation time, $t_{eq}$, is the time at which $\epsilon(t)<10^{-3}$ and remains below this limit forever. We will perform numerical calculation for this time scale later on.

\section{Numerical results}\label{2111}

We present our results from a numerical calculation of HSC at the metric \eqref{metr1}, for both the thermal and electromagnetic quench. It seems that our holographic calculation can show two distinct concepts of complexity: how much information it takes to specify a state and how many operations are required to reach the state or in other words, how long it takes to reach the state. A given state can be simple or complex based on these two categories \cite{co1}. We plot the final value of HSC, $C_{eq}$ meaning $C$ at $t\rightarrow \infty$, and relaxation time, $t_{eq}$, the time it takes for HSC to relax, in terms of the final temperature and magnetic field to illustrate the two concepts of complexity. On the one hand, the increase or decrease of $C_{eq}$, due to the amount of information needed to specify the final state by increasing temperature or magnetic field, can indicate the first category. On the other hand, the increase or decrease of $t_{eq}$ characterizes the second concept.
\begin{figure}[ht]
\begin{center}
\includegraphics[width=75 mm]{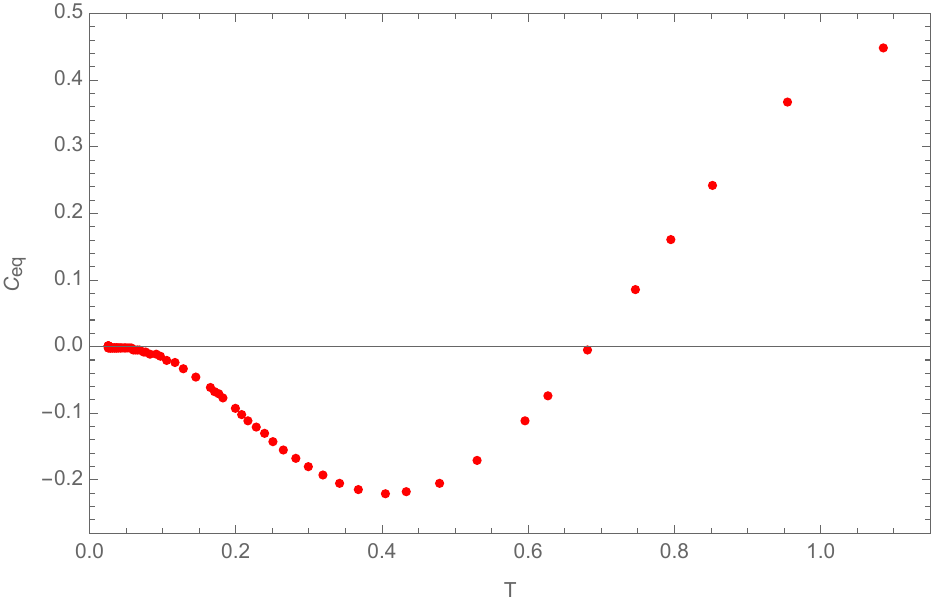}  
\caption{$C_{eq}$ as a function of $T$ for $v_{0}=0.01$ and $l=1.4$.}
\label{fig1}
\end{center}
\end{figure}

\begin{figure}[ht]
\centering 
\includegraphics[width=75 mm]{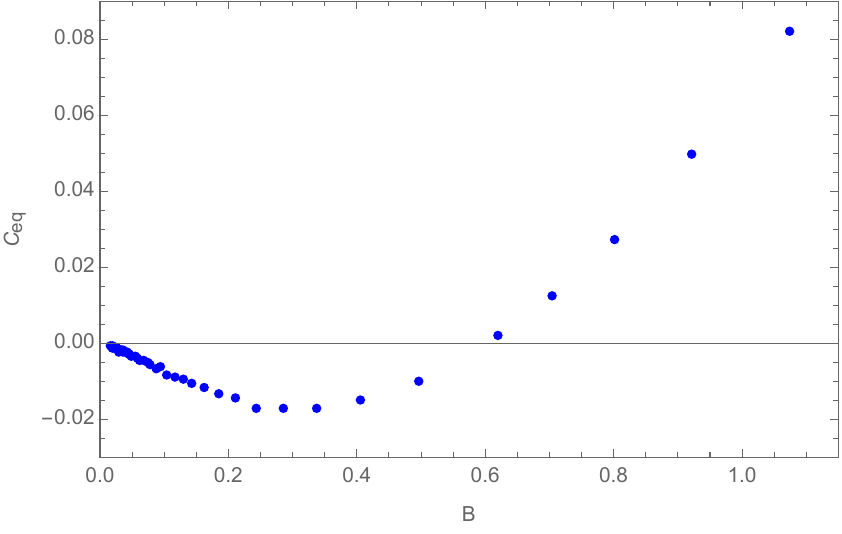} 
\caption{$C_{eq}$ in terms of $B$ for $v_{0}=0.01$ and $l=1.4$.}
\label{fig2}
\end{figure}
In figure \ref{fig1}, you can see the behavior of $C_{eq}$ as a function of the final temperature. In agreement with our previous argument \cite{co2,co3,co4}, we expect that $C_{eq}$ decrease with the increase of temperature because with the rise of temperature the number of microstates corresponding to the mixed macrostate we considered increases. At zero temperature and entropy, a unique configuration corresponding to one microstate requires more information to be specified but with the increase of the number of the microstates it is not necessary to know the details of the underlying system to specify the macrostate. Therefore, the higher the final temperature, the less information required. However, what we must notice is that this observation depends on the size of the probe, $l$. As shown in figure \ref{fig1}, $C_{eq}$ starts to increase from the temperature of $T\sim E_{l}(\equiv \frac{1}{l})$, which is when the energy of the probe, $E_{l}$, is comparable to the final temperature. In the limit of $T\gg E_{l}$ which means, large values of $T$ with fixed $E_{l}$ or small values of $E_{l}$ with fixed $T$, we expect that the degrees of freedom at widely separated scales are largely decoupled from each other. As a result, the HSC probes practically the zero temperature limit of the field theory, and thus $C_{eq}$ increases, by decreasing $E_{l}$ or increasing $T$.

When a magnetic field is applied, the magnetic moments can lower their entropy by becoming magnetized and thus with the increase of magnetic field, the entropy tends to decrease. However, in some cases, such as in a QGP plasma, the thermodynamic variables like entropy increase with the magnetic field \cite{entropyb} and references therein. Considering that we are also examining QGP-like systems, we expect that $C_{eq}$ would decrease with the increase of magnetic field, since entropy or the number of microstates corresponding to the mixed macrostate increases. This behavior is shown in figure \ref{fig2}. However, similar to the thermal case, there is a stage where $C_{eq}$ increases because $E_{l}$ is comparable to $\sqrt{B}$. In fact, when $\sqrt{B}\gg E_{l}$, our probe realizes the $B\rightarrow 0$ limit corresponding to a unique configuration or zero entropy, and hence leads to higher complexity. 

\begin{figure}[ht]
\centering 
\includegraphics[width=75 mm]{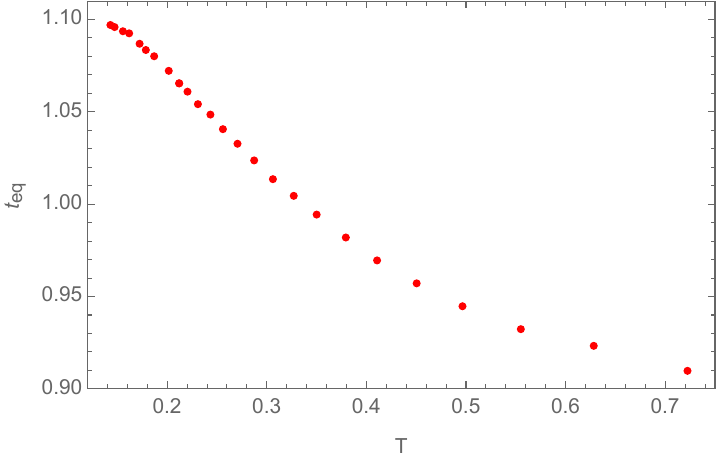}  
\caption{$t_{eq}$ as a function of $T$ for $v_{0}=0.01$ and $l=1.4$.}
\label{fig3}
\end{figure}

\begin{figure}[ht]
\centering 
\includegraphics[width=75 mm]{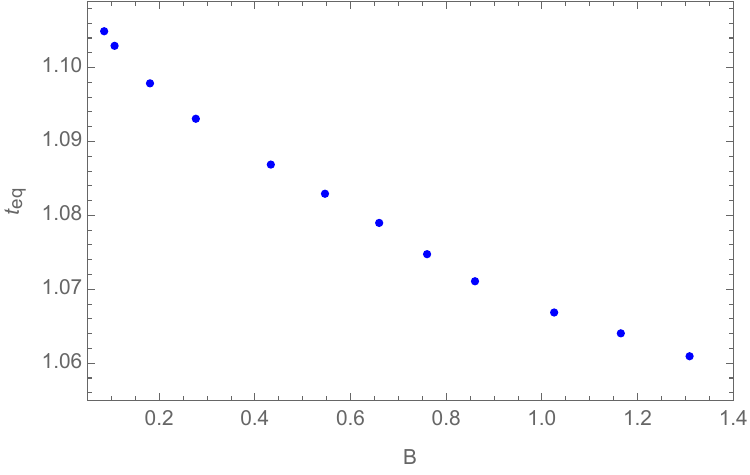} 
\caption{$t_{eq}$ as a function of $B$ for $v_{0}=0.01$ and $l=1.4$.}
\label{fig4}
\end{figure}

We plot $t_{eq}$ as a function of the final temperature, figure \ref{fig3}, and in terms of the final magnetic field, figure \ref{fig4}. In both cases, $t_{eq}$ decreases with the increase of temperature and magnetic field. When $T\geqslant E_{l}$ ($\sqrt{B}\geqslant E_{l}$) our probe investigates the state with a length scale $l$, associated with $E_{l}$,  which is larger than the intrinsic length of the theory, $l_{T}\equiv \frac{1}{T}$ ($l_{B}\equiv \frac{1}{\sqrt{B}}$). It means that the precision of the probe is not enough to study the scales smaller than $l_{T}$ ($l_{B}$). Therefore the probe, in the limit of $T (\sqrt{B})\rightarrow \infty \gg E_{l}$, reports that the temperature (magnetic field) is equal to $T$($B$) and thus in this limit in which the scales of $T$ ($\sqrt{B}$) and $E_{l}$ are widely separated, $t_{eq}\rightarrow 0$. As a result, by increasing temperature (magnetic field) for fixed value of $E_{l}$, $t_{eq}$ decreases. To check our argument, we consider the opposite limit. In the limit of $T (\sqrt{B}) \rightarrow 0 \ll E_{l}$ or equivalently $l\ll l_{T} (l_{B})$, as the probe examines the system with more precision, $t_{eq}$ increases in agreement with our argument.
\begin{figure}[ht]
\centering 
\includegraphics[width=75 mm]{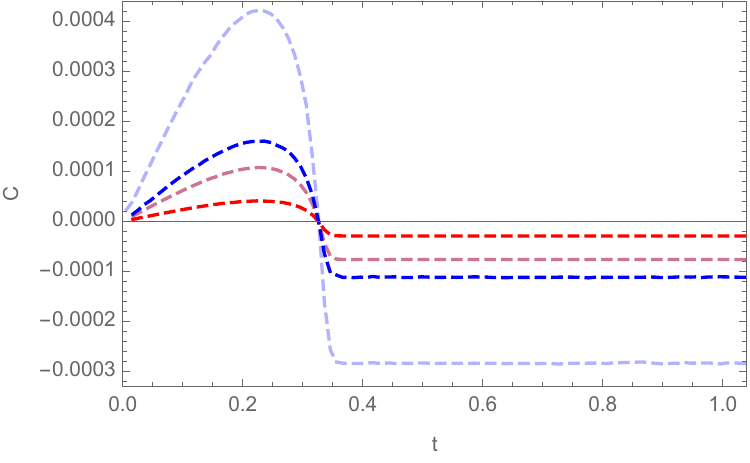}
\includegraphics[width=75 mm]{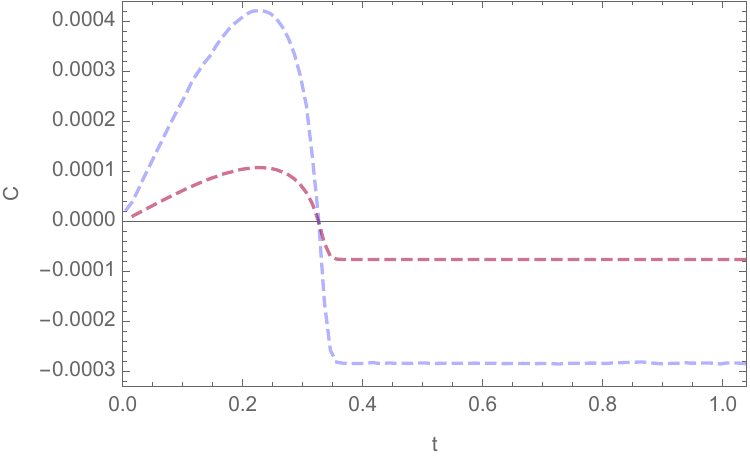}
\includegraphics[width=75 mm]{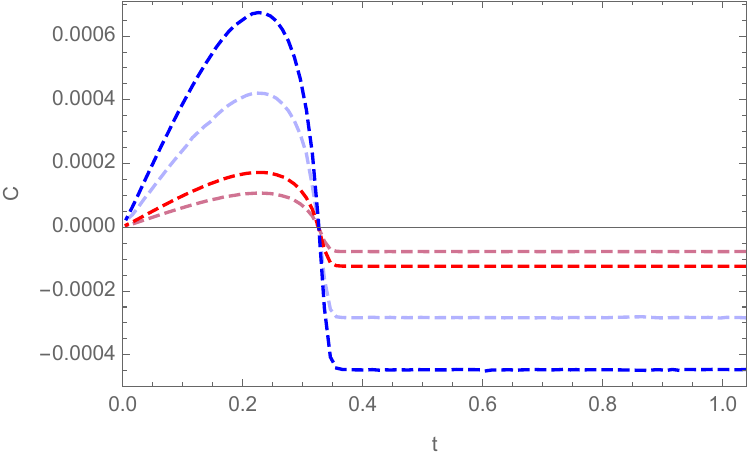}   
\caption{$C$ as a function of $t$ for, top: $B<T$, middle: $B=T$ and bottom: $B>T$. Red curves and blue curves show the behavior of $C$ for thermal and electromagnetic quenches, respectively. In three panels $l=0.5$ and $v_{0}=0.01$.}
\label{fig5}
\end{figure} 

In order to compare two different kinds of quenches, we plot $C$ as a function of time for thermal quench (red curves) and electromagnetic case (blue curves) in figure \ref{fig5}. By introducing $C_{T}$ and $C_{B}$ as $C_{eq}$ in thermal and electromagnetic quenches respectively, we list the results as follows
\begin{itemize}
\item
$B=T:$ The final temperature in thermal quench is chosen to be equal to the final magnetic field in electromagnetic quench. As shown in the middle of figure \ref{fig5}, we observe that $C_{B}<C_{T}$. 

\item 
$B<T:$ The final magnetic field is chosen to be smaller than the final temperature. As you can see in the top of figure \ref{fig5}, we observe that $C_{B}<C_{T}$ and $|C_{B}-C_{T}|$ is smaller than in the case of $B=T$.

\item 
$B>T:$ The final magnetic field is chosen to be larger than the final temperature. As shown in the bottom of figure \ref{fig5}, we observe that $C_{B}<C_{T}$ and $|C_{B}-C_{T}|$ is larger than in the case of $B=T$.
\end{itemize}

In the second and third categories, the temperature and magnetic field differences have been chosen the same and in all categories, we observe that $C_{B}<C_{T}$. As we explained earlier, both an increase in temperature in the thermal system and an increase in magnetic field in the electromagnetic system at zero temperature increase the entropy or equivalently increase the number of microstates and thus reduce the complexity. The reason why $C_{B}<C_{T}$ might be intuitively related to the fact that a thermal system has statistical fluctuations and it is a statistical system whereas an electromagnetic system at zero temperature is deterministic and predictable. Note that reliability of this statement depends on $E_{l}$. We can highlight it by considering $B\gg T$ or $T \gg B$ limits, as we will see in the next paragraph. 
\begin{figure}[ht]
\centering 
\includegraphics[width=75 mm]{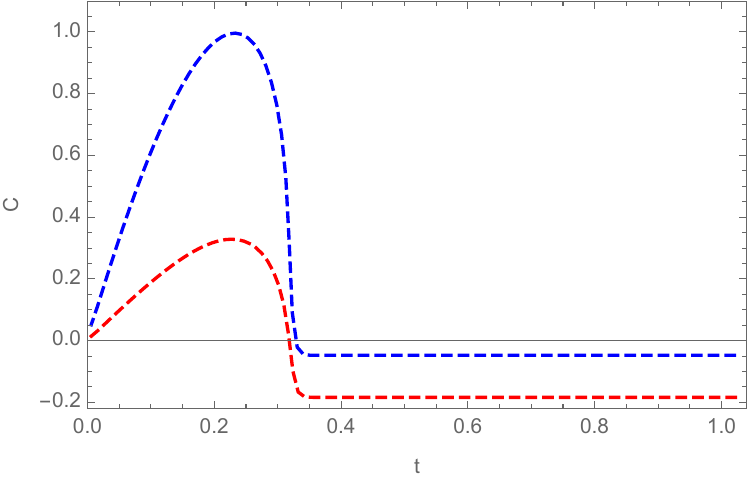}  
\includegraphics[width=75 mm]{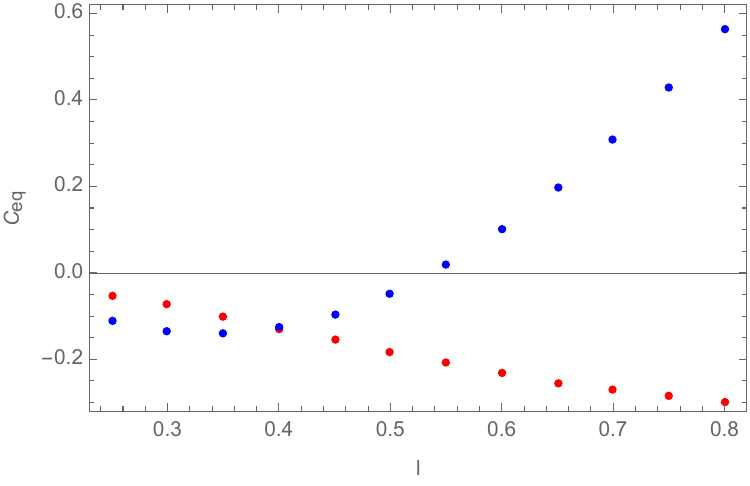} 
\caption{Top: $C$ as a function of $t$ for $B\gg T$, $l=0.5$ and $v_{0}=0.01$. Red curve indicates thermal quench and blue one is for electromagnetic quench. Bottom: $C_{eq}$ in terms of $l$ for thermal quench (red) and electromagnetic quench (blue), for $B\gg T$ and $v_{0}=0.01$.}
\label{fig6}
\end{figure} 

In the top of figure \ref{fig6}, we plotted $C$ in terms of time for $B\gg T$. In this case, as shown in the bottom of figure \ref{fig6}, whether $C_{B}$ is smaller or larger than $C_{T}$ depends on the length of the subregion, $l$. For small enough values of $l$, $C_{B}<C_{T}$. However, for large enough $l$ or in other words, small enough energy of the probe, $E_{l}$, $C_{B}>C_{T}$. 
This behavior is consistent with the results shown in figure \ref{fig2}, where $C_{B}$ starts to increase for large $B$ or small $E_{l}$, as we explained before.
$C_{T}$ also starts to increase with the increase of $l$, but it happens at larger $l$ values than in the electromagnetic case because $B\gg T$. According to this explanation, we can predict the case of $T\gg B$ as well, in which for small enough values of $l$, $C_{T}<C_{B}$ and then for large values of $l$, $C_{B}<C_{T}$. However, due to the limitations of numerical calculations, we have not plotted it.

\begin{figure}[ht]
\centering 
\includegraphics[width=75 mm]{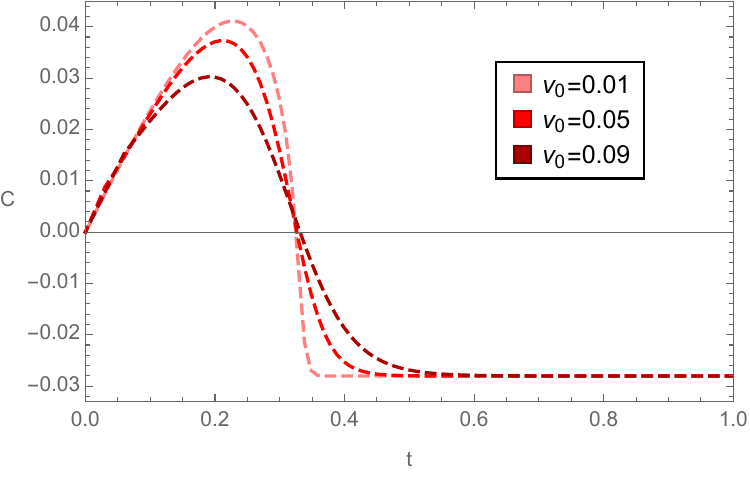}
\includegraphics[width=75 mm]{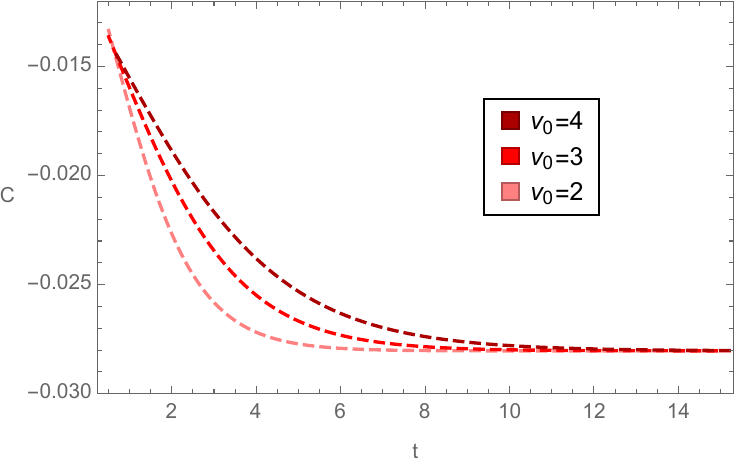}  
\caption{$C$ as a function of $t$ at $l=0.5$ and $T=0.238$ for three values of $v_{0}$ in the range of fast quenches, top, and slow quenches, bottom.}
\label{fig7}
\end{figure}

\begin{figure}[ht]
\centering 
\includegraphics[width=75 mm]{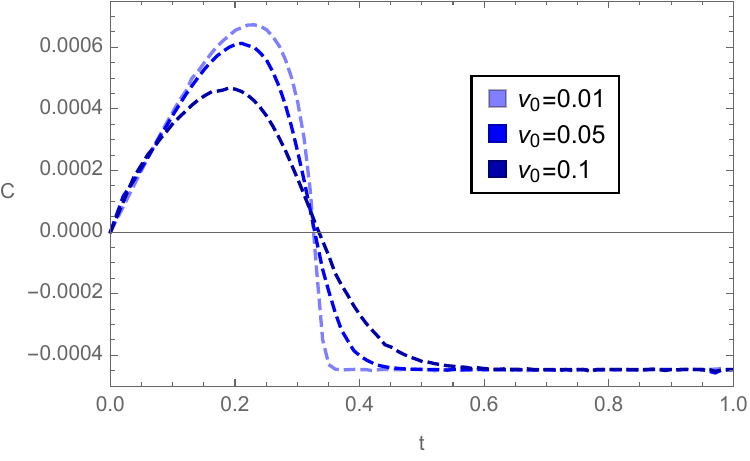}
\includegraphics[width=75 mm]{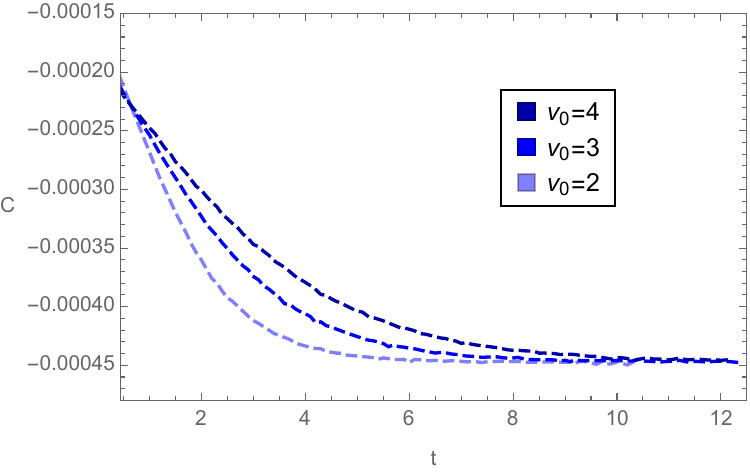}  
\caption{$C$ as a function of $t$ at $l=0.5$ and $B=0.045$ for three values of $v_{0}$ in the range of fast quenches, top, and slow quenches, bottom.}
\label{fig8}
\end{figure}
 
To study the effect of the parameter $v_{0}$, which is considered as the speed of the quench in the boundary field theory, on the evolution of the HSC, we plot $C$ in terms of time for three values of $v_{0}$ in both fast and slow quenches for thermal quench in figure \ref{fig7} and electromagnetic case in figure \ref{fig8}. In a fast quench, the system undergoes a rapid altering, while in a  slow quench, the system undergoes an adiabatic transition and has enough time to equilibrate during the evolution. In a fast quench, the HSC experiences an increase at the early stage, reaches a maximum value, and then decreases to $C_{eq}$ at late time. As you can see in both figures, different manners of energy injection into the system result in different responses during early time interval. Hence, by knowing only about $T$ or $B$ in the boundary field theory, the HSC can distinguish between fast and slow quenches. We have seen this behavior before for another different model \cite{co2}. Furthermore, the faster the energy injection, the earlier the system reaches HSC equilibrium. Therefore, in a fast quench, the system reaches equilibrium earlier than a slow quench for both thermal and electromagnetic quenches. The top panel of figure \ref{fig7} and figure \ref{fig8} shows that for thermal and electromagnetic quenches, respectively, with the increase of $v_{0}$ in the range of fast quenches, the maximum value of HSC becomes smaller, is reached sooner, and the system achieves HSC equilibrium later. The bottom panel of figure \ref{fig7} and figure \ref{fig8} shows that for thermal and electromagnetic quenches, respectively, with the increase of $v_{0}$ in the range of slow quenches, the system reaches HSC equilibrium later.

\section*{Acknowledgements}
This work is based upon research funded by Iran National Science Foundation (INSF) under Project No. 98013297.

\end{document}